\documentclass[epj]{webofc}
\usepackage[varg]{txfonts}   
\woctitle{MESON2014 - the 13$^\textrm{th}$ International Workshop on Meson Production, Properties and Interaction}
\newcommand{\cc}{\mbox{COSY--11}}
\usepackage[sort&compress]{natbib}
\begin{document}
\selectlanguage{english}
\title{Close to threshold $\eta'$ meson production in proton-proton collisions at \cc}
\author{E.~Czerwi\'nski\inst{1}\fnsep\thanks{\email{eryk.czerwinski@uj.edu.pl}} \and
        P.~Moskal\inst{1,2} \and
        M.~Silarski\inst{1}
       \\for the COSY-11 Collaboration
}
\institute{Institute of Physics, Jagiellonian University, PL-30-059 Cracow, Poland
           \and
           Institute for Nuclear Physics, Research Center J{\"u}lich, D-52425 J{\"u}lich, Germany
}
\abstract{%
We summarize results from measurements of excitation function for $pp\to pp\eta'$ reaction
performed at \cc\ detector.
The determined cross sections enabled an extraction of the scattering length of the $\eta'$-proton interaction in the vacuum. 
}
\maketitle
%
\section{Introduction}
Intensive quest for the $\eta$ and $\eta'$ bound states is currently ongoing at both
theoretical e.g.~\cite{wilkin2,bass,Mesic-Bass,EtaMesic-Hirenzaki,eta-prime-mesic-Nagahiro-Oset,
ETA-Friedman-Gal,Wycech-Acta,Mesic-Kelkar}
and experimental levels e.g.
  COSY~\cite{COSY11-MoskalSmyrski,WASA-at-COSY-SkuMosKrze,Adlarson2013,GEM,3Heeta-PL-Smyrski,3Heeta-Mersmann},
  ELSA~\cite{cbelsa},
  GSI~\cite{eta-prime-mesic-GSI-tanaka,eta-prime-mesic-GSI-Itahashi},
  JINR~\cite{JINR},
  JPARC~\cite{eta-mesic-JPARC-Fujioka,eta-mesic-JPARC-Fujioka-Itahashi},
  LPI~\cite{LPI}, and
  MAMI~\cite{ELSA-MAMI-plan-Krusche,gamma3He-Pheron}.
These studies were already supported by data provided by the \cc\ collaboration~\cite{Moskal1,Moskal2,Khoukaz,pklaja}
including determination of the total width of the $\eta'$ meson~\cite{ErykPhD,eryk_prl}.
In addition, the first rough estimation of the $\eta'$-$N$ interaction from the excitation function of the cross section for the $pp\to pp\eta'$ reaction
was also performed~\cite{Swave}. Recent precise measurement in this field from the \cc\ experiment~\cite{eryk_prl2} allows to summarize results on
the $\eta'$ meson production cross section in proton-proton collisions at \cc.
\section{Motivation}
It is impossible to prepare a beam or target out of short lived particles like the $\eta$ or $\eta^{\prime}$ mesons.
Therefore their interaction with other hadrons cannot be investigated in the standard way via scattering experiments.
However, production of these mesons close to the kinematical threshold with low relative velocities with respect to nucleons
gives a chance to study their interaction with nucleons.
It may manifest itself as structures in a meson-nucleon invariant mass distributions
and as enhancement in the excitation function with respect
to predictions based on the assumption that the kinematically available phase space is  homogeneously populated.

Measurements of the $\eta-$ and $\eta'-$nucleon and nucleus systems
may yield valuable new information about dynamical chiral and axial U(1) symmetry breaking in low energy QCD.
The binding energies, meson-nucleon scattering lengths and in-medium masses of the $\eta$ and $\eta'$
are sensitive to the flavour-singlet component in the mesons and hence to
the non-perturbative glue associated with axial U(1) dynamics~\cite{bass,Mesic-Bass}.
QCD inspired models including confinement, chiral and axial U(1) dynamics yield a range of predictions for
the $\eta$ and $\eta'$ nucleon scattering lengths and binding in nuclei.

The quark condensate is modified in nucleus which changes the properties of hadrons in nuclear medium and
these medium modifications can be understood at the quark level through coupling of the scalar isoscalar $\sigma$ 
(and also $\omega$ and $\rho$) mean fields in the nucleus to the light quarks in the hadron.
\section{The COSY--11 experiment}
The collision of a proton from the COSY stochastically
cooled beam~\cite{Maier} with a hydrogen cluster target proton of \cc~\cite{Brauksiepe} may cause an $\eta'$ meson creation.
The ejected protons of the $pp\to pp\eta'$ reaction are then separated from the circulating beam
by the magnetic field due to their lower momenta and were registered by the detection system consisting of drift chambers and scintillation
counters.
The reconstruction of the momentum vector for each registered particle is based on
the measurement of track direction by means of the drift chambers and
the knowledge of dipole magnetic field.
Together with the independent determination of the particle velocity from the
measured time of flight between scintillator detectors the particle identification is provided.
Knowledge of the momenta of both protons before and after the reaction allows to calculate the mass of unobserved particles.
Number of reconstructed $\eta'$ mesons together with luminosity determination based on the cross section for elastically scattered $pp$ events and
registered number of $pp\to pp$ events allows for $pp\to pp\eta'$ cross section determination.

\begin{figure}[h]
\centering
    \includegraphics[width=0.47\textwidth]{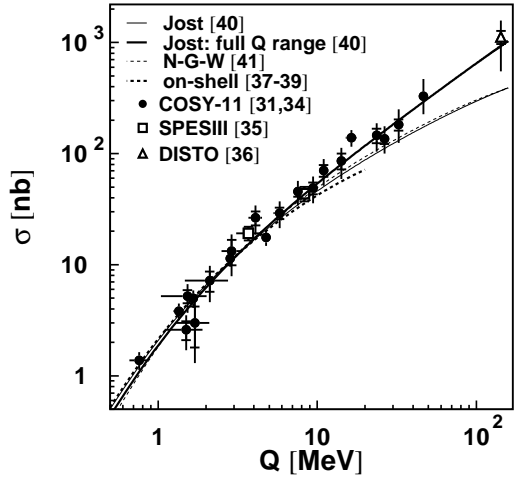}
 \caption{
 The total cross sections for the $pp\to pp\eta'$ reaction as a function of the excess energy $Q$.
 Experimental data with the statistical and systematic errors separated by dashes
 are marked as solid circles for the COSY--11 experiment~\cite{eryk_acta,eryk_prl2},
 as open squares for SPESIII measurements ~\cite{Hibou} and as open triangle for the DISTO experiment~\cite{Balestra}.
 In addition the superimposed lines show results of fits 
 parameterizing the $pp$-FSI
 enhancement factor as in
 Refs.~\cite{noyes995,noyes465,naisse506} (thick dashed line),
 inverse of the squared Jost function~\cite{Druzhinin} (thin solid line)
 and Niskanen-Goldberger-Watson model~\cite{Shyam} (thin dashed line)
 with the $\eta'$-proton scattering length as a free parameter.
 The thick dashed line is shown only in the range of applicability of
 the formula used for the enhancement factor~\cite{noyes995}.
 For comparison the thick solid line shows result of the fit obtained for the whole $Q$ range with $pp$-FSI parametrization
 from Ref.~\cite{Druzhinin}.
 \label{c11fig}
         }
\end{figure}

Measurements of the total cross section for $pp\to pp\eta'$ reaction together with theoretical excitation functions are summarized in Figure~\ref{c11fig}.
\cc\ data are gathered in Table~\ref{c11tab}.
\begin{table}[h]
\centering
  \caption{Production cross-sections for the $\eta'$ meson
	in proton-proton collisions measured at \cc\ detector~\cite{eryk_acta,eryk_prl2}.
        Excess energy $Q$ is given with the absolute systematic uncertainty and the cross sections values are given with the statistical and systematical uncertainty,
        respectively.}
\begin{tabular}{ r@{.}l c r@{.}l | r l l}
  \multicolumn{5}{c}{$Q~\left[MeV\right]$} &
  \multicolumn{3}{c}{$\sigma_{pp\to pp\eta'}~\left[nb\right]$} \\
  \noalign{\smallskip}
  \hline
  \hline
  \noalign{\smallskip}
 0&76&$\pm$&0&10&  1.38&$\pm$~0.08&$\pm$~0.17 \\
 1&35&$\pm$&0&10&  3.82&$\pm$~0.19&$\pm$~0.47 \\
 1&5 &$\pm$&0&4 &  2.6 &$\pm$~0.5 &$\pm$~0.4 \\
 1&53&$\pm$&0&49&  5.2 &$\pm$~0.7 &$\pm$~0.8 \\
 1&66&$\pm$&0&10&  4.97&$\pm$~0.28&$\pm$~0.61 \\
 1&7 &$\pm$&0&4 &  3.0 &$\pm$~1.2 &$\pm$~0.5 \\
 2&11&$\pm$&0&64&  7.2 &$\pm$~1.5 &$\pm$~1.1 \\
 2&84&$\pm$&0&10& 11.41&$\pm$~0.40&$\pm$~1.39 \\
 2&9 &$\pm$&0&4 & 13.3 &$\pm$~3.4 &$\pm$~2.0 \\
 4&1 &$\pm$&0&4 & 26.4 &$\pm$~3.8 &$\pm$~4.0 \\
 4&78&$\pm$&0&10& 17.58&$\pm$~0.64&$\pm$~2.15 \\
 5&80&$\pm$&0&50& 29.2 &$\pm$~3.5 &$\pm$~4.4 \\
 7&57&$\pm$&0&51& 45.5 &$\pm$~4.5 &$\pm$~6.8 \\
 9&42&$\pm$&0&53& 49.0 &$\pm$~5.9 &$\pm$~7.4 \\
10&98&$\pm$&0&56& 70.5 &$\pm$~8.6 &$\pm$~11\\
14&21&$\pm$&0&57& 86   &$\pm$~14  &$\pm$~13\\
16&4 &$\pm$&1&3 &139   &$\pm$~3   &$\pm$~21\\
23&64&$\pm$&0&64&146   &$\pm$~20  &$\pm$~22\\
26&5 &$\pm$&1&0 &136   &$\pm$~14  &$^{+22}_{-26}$ \\
32&5 &$\pm$&1&0 &182   &$\pm$~21  &$^{+36}_{-48}$ \\
46&6 &$\pm$&1&0 &329   &$\pm$~18  &$^{+85}_{-122}$
\end{tabular}
  \label{c11tab}
\end{table}
\section{Conclusions}
The $\eta'$ production cross sections in proton-proton collisions provided by \cc\ collaboration for the last 16 years~\cite{Moskal1,Moskal2,Khoukaz,pklaja,eryk_prl2}
together with the recent precise determination of the 
$\eta'$-proton scattering length in free space~\cite{eryk_prl2}
constitute a significant contribution to the study of the $\eta'$ properties
and the search of $\eta'$ bound state~\cite{moskal_few}.
\begin{acknowledgement}
We acknowledge support
by the Polish National Science Center through grants No. 2011/03/B/ST2/01847, 2011/01/B/ST2/00431, 
by the FFE grants of the Research Center J\"{u}lich,
the European Commission through European Community-Research Infrastructure Activity under FP6 project
Hadron Physics (contract number RII3-CT-2004-506078),
and by the Polish Ministry of Science and Higher Education through grant No. 393/E-338/STYP/8/2013.
\end{acknowledgement}

\end{document}